\documentclass[12pt]{iopart}

\usepackage{booktabs}
\usepackage[colorlinks = true,
            linkcolor = cyan,
            urlcolor  = blue,
            citecolor = blue,
            anchorcolor = black]{hyperref}	
\usepackage{soul} 
\usepackage{setspace}
\usepackage{amssymb}
\usepackage{graphicx}
\usepackage[labelsep=period]{caption}
\usepackage{float}

\usepackage[dvipsnames]{xcolor}

\usepackage{soul}

\makeatletter
\newcommand{\smallsym}[2]{#1{\mathpalette\make@small@sym{#2}}}
\newcommand{\make@small@sym}[2]{%
  \vcenter{\hbox{$\m@th\downgrade@style#1#2$}}%
}
\newcommand{\downgrade@style}[1]{%
  \ifx#1\displaystyle\scriptstyle\else
    \ifx#1\textstyle\scriptstyle\else
      \scriptscriptstyle
  \fi\fi
}
\makeatother

\begin{document}

\title{MHD simulation of tilt instability during the dynamic FRC magnetic compression process}

\author{Yiming Ma$^1$, Ping Zhu$^{1,2*}$, Bo Rao$^{1*}$ and Haolong Li$^3$}

\address{
$1$ State Key Laboratory of Advanced Electromagnetic Technology, International Joint Research Laboratory of Magnetic Confinement Fusion and Plasma Physics, School of Electrical and Electronic Engineering, Huazhong University of Science and Technology, Wuhan, 430074, China
\\
$2$ Department of Nuclear Engineering and Engineering Physics, University of Wisconsin-Madison, Madison, Wisconsin 53706, USA \\
$3$ Department of Physics, School of Science, Tianjin University of Science and Technology, Tianjin 300457, China

}
\ead{zhup@hust.edu.cn, borao@hust.edu.cn}

\vspace{10pt}


\begin{abstract}

The nonlinear evolution of the tilt instability in a field reversed configuration (FRC) during the dynamic magnetic compression process has been investigated using magnetohydrodynamic (MHD) simulations with the NIMROD code [C. R. Sovinec \textit{et al.}, J. Comput. Phys. \textbf{195}, 355 (2004)]. The tilt mode induces significant deformations in the linear growth phase and results in complete confinement loss of the FRC in the nonlinear phase, with no evidence of dynamic nonlinear stabilization. The growth rate of the tilt mode increases with the compression field ramping rate  and approaches an asymptotic value. Toroidal flow can reduce both the growth rate and the nonlinear saturation amplitude of the tilt mode. The stabilizing effect of the toroidal rotation is enhanced with higher compression field ramping rates due to the spontaneous toroidal field generation and increased flow shear during compression. Although the tilt mode remains unstable with a toroidal rotation Mach number close to 0.5, the onset of tilt distortion can be delayed, allowing a magnetic compression ratio up to 5.3 before the compressional heating terminates.

\end{abstract}

\vspace{0.5pc}
\clearpage

\section{Introduction}

The field reversed configuration (FRC) is a compact toroid primarily sustained by the poloidal magnetic field~\cite{tuszewski1988field}.  Due to its attractive properties such as high plasma $\beta$, linear device geometry, natural divertors and ease of axial translation~\cite{rej1986experimental, steinhauer1996frc, liao2022first, liao2022dynamic, sekiguchi2018super, kobayashi2021experimental, asai2021observation, asai2019collisional}, the FRC is recognized as a promising alternative path for compact fusion reactors as well as advanced fusion fuel applications~\cite{steinhauer2011review, gota2024enhanced, kirtley2023fundamental}. The fusion FRC plasma can either be achieved by pulsed compression~\cite{spencer1983adiabatic, woodruff2008adiabatic, ma2023mhd, degnan2013recent}, or by sustaining a steady FRC with additional heating~\cite{asai2019collisional, okada2001experiments, gota2021overview, galea2023princeton, hoffman2009advances, ono2024ion}. The pulsed magnetic compression has been demonstrated as an effective heating method for FRCs in experiments~\cite{kirtley2023fundamental, rej1992high, slough2011creation}. However, gross magnetohydrodynamic (MHD) instabilities remain a critical challenge for the pulsed FRC system~\cite{turchi2025progress, hill2008report}, with the potential of tilt instability being the most dangerous. 

The tilt instability is initially identified through analytical studies on the spheromak~\cite{rosenbluth1979mhd}, and subsequent MHD analyses also find the FRCs susceptible~\cite{clemente1981tilting, hammer1981mhd}. The antiparallel current configuration between the FRC and external coils can drive a tilt motion to reduce the potential energy of the FRC~\cite{bellan2000spheromaks}. Since the tilt mode has been shown a global internal mode in elongated FRCs, the external stabilization methods are ineffective and strong kinetic effects may be required for its stabilization~\cite{belova2000numerical, milroy1989nonlinear}. Moreover, MHD analyses indicate that no equilibrium profile is inherently stable to the tilt mode~\cite{schwarzmeier1983magnetohydrodynamic,cobb1993profile, iwasawa2000ideal}. The toroidal flow has modest effects on the tilt instability for realistic rotation speed~\cite{belova2000numerical, milroy1989nonlinear, horiuchi1989full}. Numerical simulations further demonstrate the tilt stability for FRCs in the low $\bar{s}$ regime ($\bar{s}<3$), where $\bar{s}$ is the averaged gyroradius between the magnetic axis and the separatrix~\cite{nishimura1997tilt, barnes1986kinetic, belova2004kinetic}, however, empirical confinement scalings suggest that practical fusion reactors need to operate within the MHD regime above a minimum $\bar{s}$ of 10-20~\cite{hill2008report, slough1995transport, slough1992confinement, belova2001numerical}. Whereas it is promising to extend the reactor potential of FRC to the large-$\bar{s}$ MHD regime in steady state~\cite{slough1992confinement, slough1993stability}, the stability of FRC during a dynamic compression against tilt mode remains unclear.

In this work, we perform nonlinear MHD simulations of the tilt mode during dynamic FRC magnetic compression using the NIMROD code~\cite{sovinec2004nonlinear}. The effects of the magnetic field ramping rate and the toroidal flow on tilt instability are also investigated. Considering that the effectiveness of active control methods is limited by the short pulse duration and rapid FRC shape changes, the evolution of the tilt mode is examined throughout the compression process with special attention to the spontaneous stabilizing effects. Whereas the FRC is confirmed unstable to the tilt instability in the MHD regime during compression, the stabilizing effects of toroidal flow are found enhanced by the more dynamic conditions due to the spontaneous toroidal field generation and the increased flow shear.

The rest of paper is organized as follows. After a brief description of the numerical model in section~\ref{sec: model}, the simulation results are presented in section~\ref{sec: results}. The discussions and conclusions will be given in section~\ref{sec: Discussion and conclusions}.

\section{Numerical model}
\label{sec: model}

The following resistive single-fluid MHD equations are solved using the NIMROD code~\cite{sovinec2004nonlinear}:

\begin{equation}
  \frac{\partial N}{\partial t} + \nabla \cdot (N\vec{u})= 0 \, ,
\end{equation}

\begin{equation}
  \rho(\frac{\partial \vec{u}}{\partial t} + \vec{u}\cdot \nabla\vec{u} ) = \vec{J}\times \vec{B} - \nabla p - \nabla \cdot \stackrel{\leftrightarrow}{\Pi} \, ,
\end{equation}

\begin{equation}
  \label{eqn: mhd temperature}
\frac{N}{\gamma-1} (\frac{\partial}{\partial t} + \vec{u}\cdot \nabla ) T = -p \nabla \cdot \vec{u} + Q \, ,
\end{equation}

\begin{equation}
  \frac{\partial \vec{B}}{\partial t} = - \nabla \times \vec{E} \, ,
\end{equation}

\begin{equation}
  \vec{E} = -\vec{u}\times \vec{B} + \eta \vec{J} \, ,
\end{equation}
where $N$ is the plasma number density, $\vec{u}$ the plasma velocity, $\vec{J}$ the current density, $p$ the pressure, $\rho$ the mass density,  $\vec{B}$ the magnetic field,  $\stackrel{\leftrightarrow}{\Pi}$ the viscosity tensor, $T=T_i+T_e$ the total temperature, $\vec{E}$ the electric field, $\eta$ the resistivity and $Q=\eta J^2$ represents resistive heating. The core Spitzer resistivity is initialized to a Lundquist number of $5\times 10^4$ based on the length scale of wall radius. An isotropic viscosity corresponding to the Reynolds number of $1\times 10^3$ and a parallel viscosity of $3\times10^3 \mathrm{m^2 / s}$ are also adopted.

In the NIMROD code, the spatial discretization is performed using finite elements in the $(r,z)$ poloidal plane along with a Fourier decomposition in the azimuthal direction in the cylindrical system of coordinates $(r,\theta,z)$. A magnetic compression boundary condition is imposed on the $r=r_w$ wall with the time ramping of the axial magnetic compression field~\cite{ma2023mhd}. The compression boundary condition is assumed axisymmetric and thus applied only to the $n=0$ Fourier component, where $n$ is the toroidal mode number.

The 3D stability calculations presented in this work are initialized with the 2D equilibria obtained by solving the Grad-Shafranov (GS) equation with flow using the NIMEQ code~\cite{li2021solving,howell2014solving}. In particular, in the presence of a toroidal flow $\vec{u} = \Omega(\psi)r \hat{e}_\theta$, the GS equation we solve takes the following form~\cite{belova2000numerical, evangelias2016axisymmetric, guazzotto2009magnetic, maschke1980exact, sen1997tokamak, li2021formation}
\begin{equation}
\label{eqn: gs}
\Delta^* \psi=  -\mu_0 r^2 \frac{\partial p}{\partial \psi}, 
\end{equation}

\begin{equation}
  \label{eqn: pres wi flow}
p(\psi, r)=p_0(\psi) \exp \left(\frac{m_i r^2 \Omega^2}{2 T}\right),
\end{equation}

\begin{equation}
    \label{eqn: rho wi flow}
\rho(\psi, r)=\rho_0(\psi) \exp \left(\frac{m_i r^2 \Omega^2}{2 T}\right),
\end{equation}
where $\Delta^*=r^2\nabla\cdot(\nabla/r^2)$, $T(\psi)$ is a flux function, $p_0(\psi)$ and $\rho_0(\psi)$ denote the static equilibrium pressure and mass density respectively. The static GS equation is recovered in the limit of $\Omega \rightarrow 0$.

The influence of different toroidal rotation profiles on the tilt instability is also investigated, including the rigid rotation (RR), a double-peaked rotation similar to that in reference~\cite{belova2000numerical} and a single-peaked rotation profile with the maximum flow speed near the O-point~\cite{ono2003spontaneous} where the $\Omega(\psi)$ is modeled using Gaussian shaped function and a cubic function for the respective cases~\cite{howell2014solving, li2021formation}.

\section{Simulation results}
\label{sec: results}

\subsection{Tilt mode evolution during dynamic compression}
\label{sec: typical case}

The simulations initialize from an $n=1$ axial velocity perturbation to the 2D FRC equilibrium. The perturbation has a peak amplitude of $0.01V_{A}$, where $V_{A}=B_w / (\mu_0 m_i n_a)^{1/2}$ is the characteristic Alfv\'{e}n velocity of the initial equilibrium, $n_a$ is the density at the magnetic axis and $B_w$ is the maximum axial magnetic field at the $z=0$ middle plane. The simulations are numerically convergent with the inclusion of the $n=0$ and $n=1$ Fourier components, since the further addition of higher-$n$ modes yields virtually the identical results. The standard simulation in this section employs a moderate compression field ramping rate of $0.12B_{w0}/\mathrm{\mu s}$ similar to that in the FRX-C/LSM (Field-Reversed Experiment-C/Large `S' Modification) experiments~\cite{rej1992high}, where the subscript $0$ refers to the initial equilibrium state. The initial equilibrium has an $x_s=r_s/r_w\sim1$ corresponding to the optimal compression condition and an elongation $E=l_s/(2r_s)\sim5$, where $r_s$ and $l_s$ are the separatrix radius and length respectively.

\begin{figure*}[!htbp]
  \centering
  \includegraphics[width=0.65\textwidth]{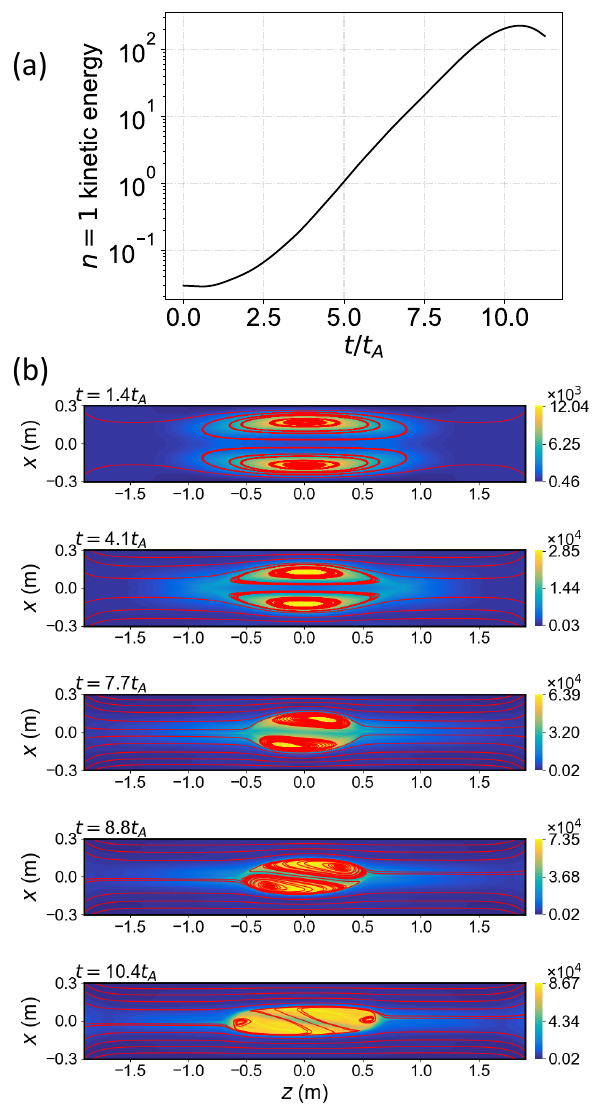}
  \caption{(a) The time history of the $n=1$ kinetic energy and the corresponding; (b)  temporal evolution of magnetic field lines (red line) and pressure contours (colored) in the poloidal plane for the case with a compression field ramping rate of $0.12B_{w0}/\mathrm{\mu s}$.}
  \label{fig: e1streamanden}
\end{figure*}

The temporal evolution of the magnetic field lines and pressure contours in the $(r,z)$ planes of $\theta=0$ and $\theta=\pi$ are shown in figure~\ref{fig: e1streamanden}, where $x=r\cos\theta$ and  $t_A=r_w/V_A$ is the characteristic Alfv\'{e}n time of the initial equilibrium. The corresponding time dependence of the $n=1$ kinetic energy exhibits a rapid growth rate $\gamma\sim 3\gamma_0$, and here $\gamma_0 = 2V_A/l_{s0}$ is the reference growth rate of the tilt mode based on previous MHD calculations~\cite{belova2000numerical, milroy1989nonlinear, belova2001numerical}. As shown in figure~\ref{fig: e1streamanden}, the FRC becomes increasingly oblate as it shrinks both radially and axially as the compression process proceeds, which is accompanied by the growing shape distortion typical of the tilt mode. The significant deformation starts during the linear growth phase and eventually leads to the FRC disruption in the nonlinear growth phase, where the closed field lines near the X-points tear open and reconnect with open field lines until the confinement is nearly lost.

\begin{figure*}[!htbp]
  \centering
  \includegraphics[width=0.8\textwidth]{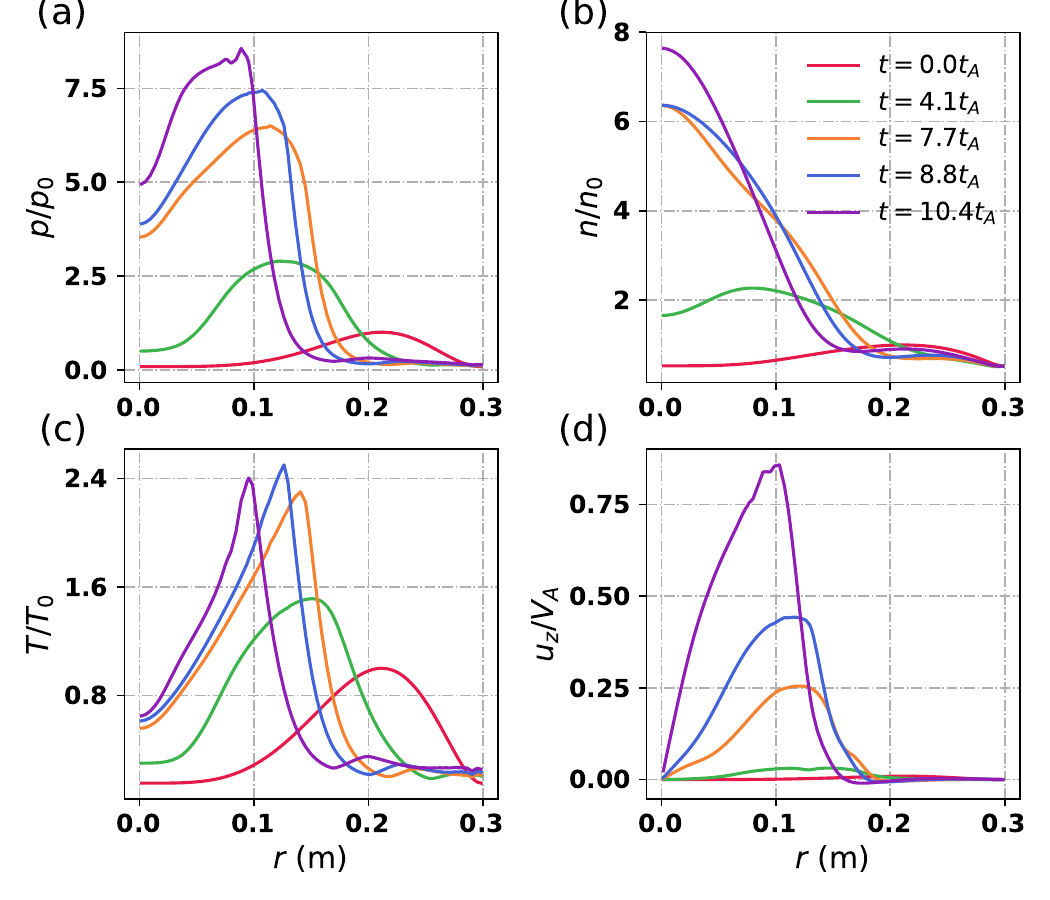}
  \caption{Radial profiles along the $z=0$ middle plane of the normalized (a) pressure, (b) density, (c) plasma temperature and (d) axial velocity at different times during compression respectively for the same case as shown in figure~\ref{fig: e1streamanden}.}
  \label{fig: e1z0slice}
\end{figure*}

The dynamic compression process can be also illustrated with  the radial profiles of the pressure, density, plasma temperature and axial velocity along the $z=0$ middle plane at different times in figure~\ref{fig: e1z0slice}. The pronounced temperature gradient can be explained by the adiabatic limit of equation~(\ref{eqn: mhd temperature}) in the absence of thermal conduction. In the nonlinear phase of the tilt mode evolution, the temperature decreases and the tilting axial velocity can reach up to $0.8V_A$. The tilt instability terminates the compressional heating in the early stage of magnetic compression when the compression ratio $B_{w1}/B_{w2}<3$.

\begin{figure*}[!htbp]
  \centering
  \includegraphics[width=1.0\textwidth]{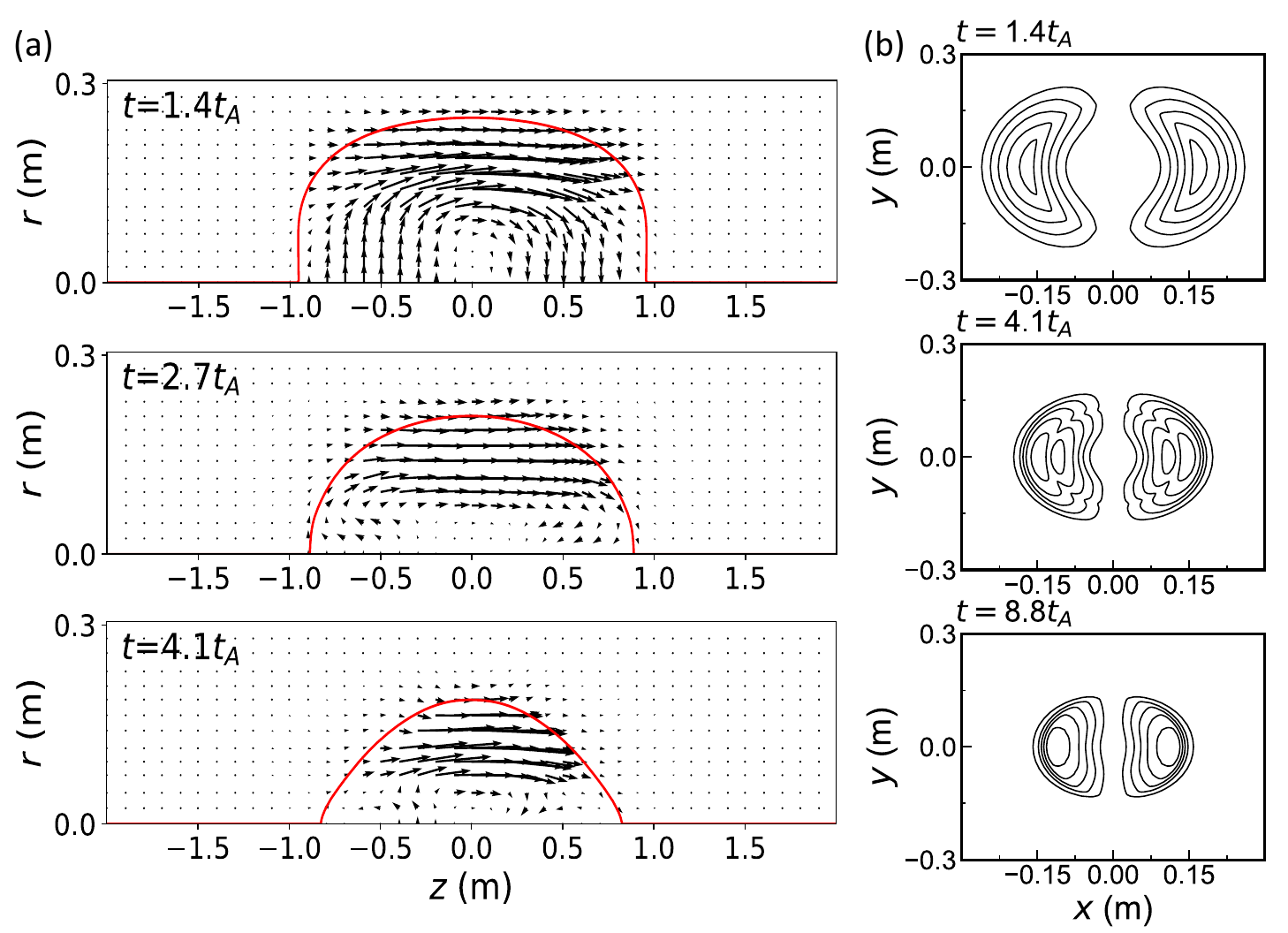}
  \caption{(a) Velocity vectors and the separatrix locations (red line) in the poloidal plane and (b) axial velocity contours in the $z=0$ middle plane of the $n=1$ tilt mode at different times during compression for the same calculation as shown in figure~\ref{fig: e1streamanden}.}
  \label{fig: e1modestruct}
\end{figure*}

The velocity vector plots reveal the tilt mode structure in the linear growth phase in figure~\ref{fig: e1modestruct}, where the red line denotes the separatrix. During the compression process, the separatrix rapidly evolves from a racetrack-like shape to a more elliptical form. The tilt mode is primarily internal, with small displacements outside the separatrix that may potentially serve as experimental indicators of the tilt mode. Additionally, the tilt mode grows locally without rotation in the $\theta$ direction.

The tilt mode is highly unstable in the typical simulation with a moderate compression field ramping rate. During the dynamic compression process, the tilt instability causes significant deformation within $10t_A$ while the FRC contracts rapidly. Consequently, active stabilization methods such as the neutral beam injection~\cite{barnes1991stabilization} may become less effective. It is thus important to find out whether simply increasing the magnetic field ramping rate can prevent the tilt deformation before peak compression, as has been demonstrated for rotational instability with sufficiently rapid magnetic compression~\cite{kirtley2023fundamental, rej1992high}. Meanwhile, spontaneous stabilization mechanisms also merit investigation for the pulsed compression. These issues will be examined in the following two sections.

\subsection{Effects of compression magnetic field ramping rate}

\begin{figure*}[!htbp]
  \centering
  \includegraphics[width=0.7\textwidth]{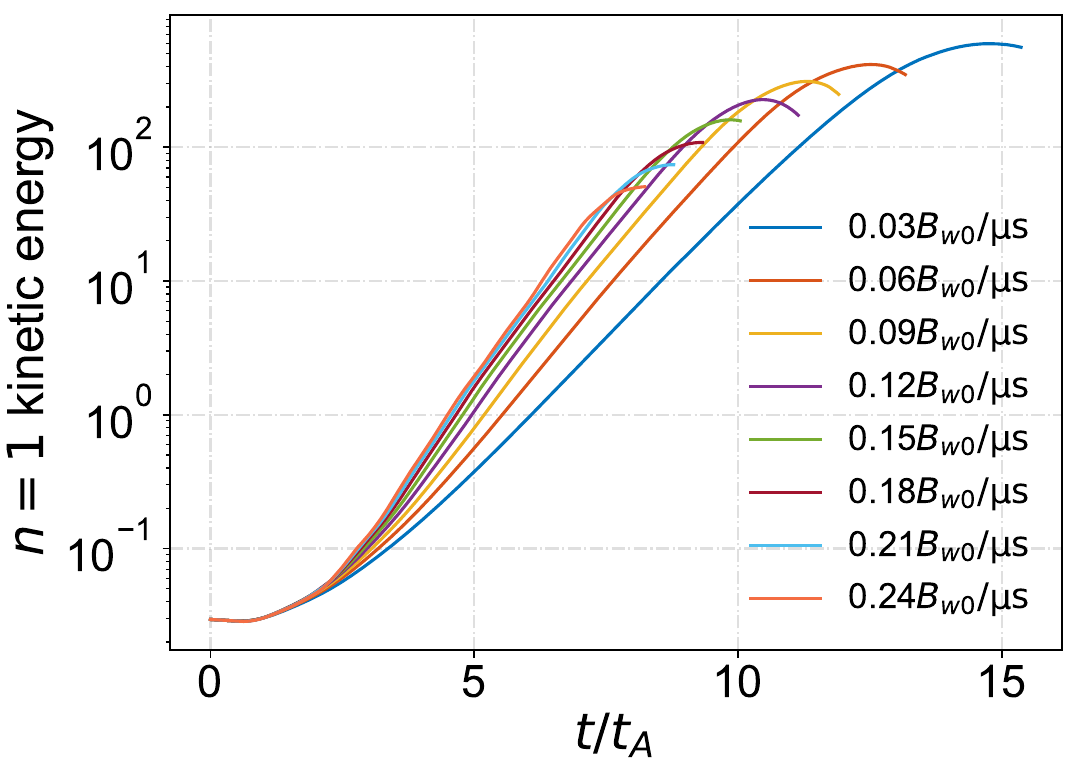}
  \caption{Temporal evolution of the $n=1$ kinetic energy for the cases with various compression field ramping rates.}
  \label{fig: rpenbin}
\end{figure*}

A higher ramping rate corresponds to more intensified dynamics, leading to significant changes in plasma parameters and deviations of the FRC from equilibrium during compression~\cite{ma2023mhd}. The previous section examined the tilt mode evolution with a moderate ramping rate $0.12B_{w0}/\mathrm{\mu s}$ that is consistent with FRX-C/LSM experiments~\cite{rej1992high}. Figure~\ref{fig: rpenbin} presents the time evolution of the $n=1$ kinetic energy for cases with compression field ramping rates ranging from $0.03B_{w0}/\mathrm{\mu s}$ to $0.24B_{w0}/\mathrm{\mu s}$. As the ramping rate increases, the nonlinear saturation magnitude of the $n=1$ kinetic energy decreases, even though the linear growth rate increases to an asymptotic value approximately 3.3 times as large.

\begin{figure*}[!htbp]
  \centering
  \includegraphics[width=1.1\textwidth]{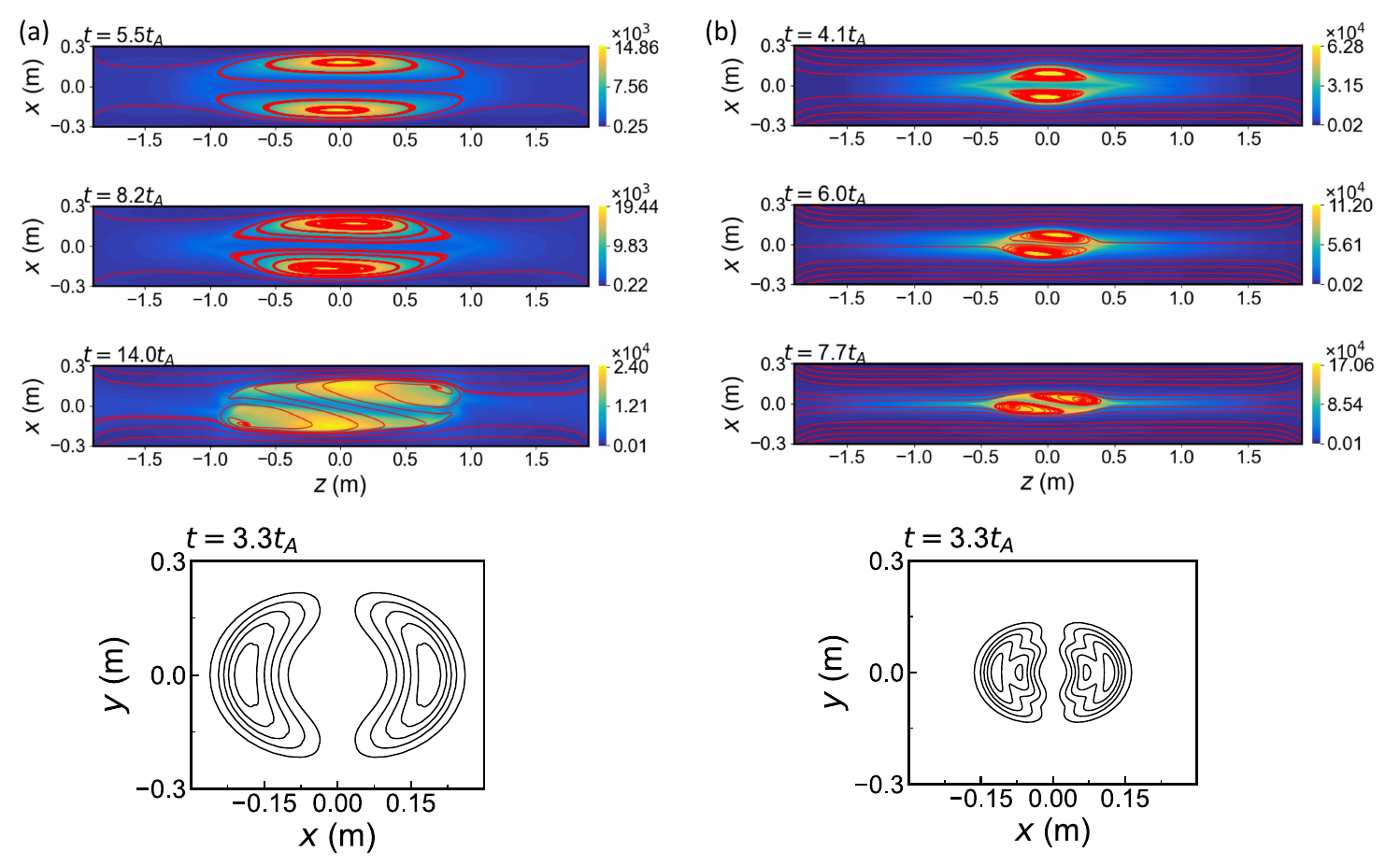}
  \caption{Temporal evolution of magnetic field lines (red line) with pressure contours (colored) in the poloidal plane (top) and the axial velocity contours of the $n=1$ tilt mode (bottom) for the cases with the compression field ramping rates of (a) $0.03B_{w0}/\mathrm{\mu s}$ and (b) $0.24 B_{w0}/\mathrm{\mu s}$.}
  \label{fig: rpstream_z0uz_comp}
\end{figure*}

\begin{figure*}[!htbp]
  \centering
  \includegraphics[width=0.7\textwidth]{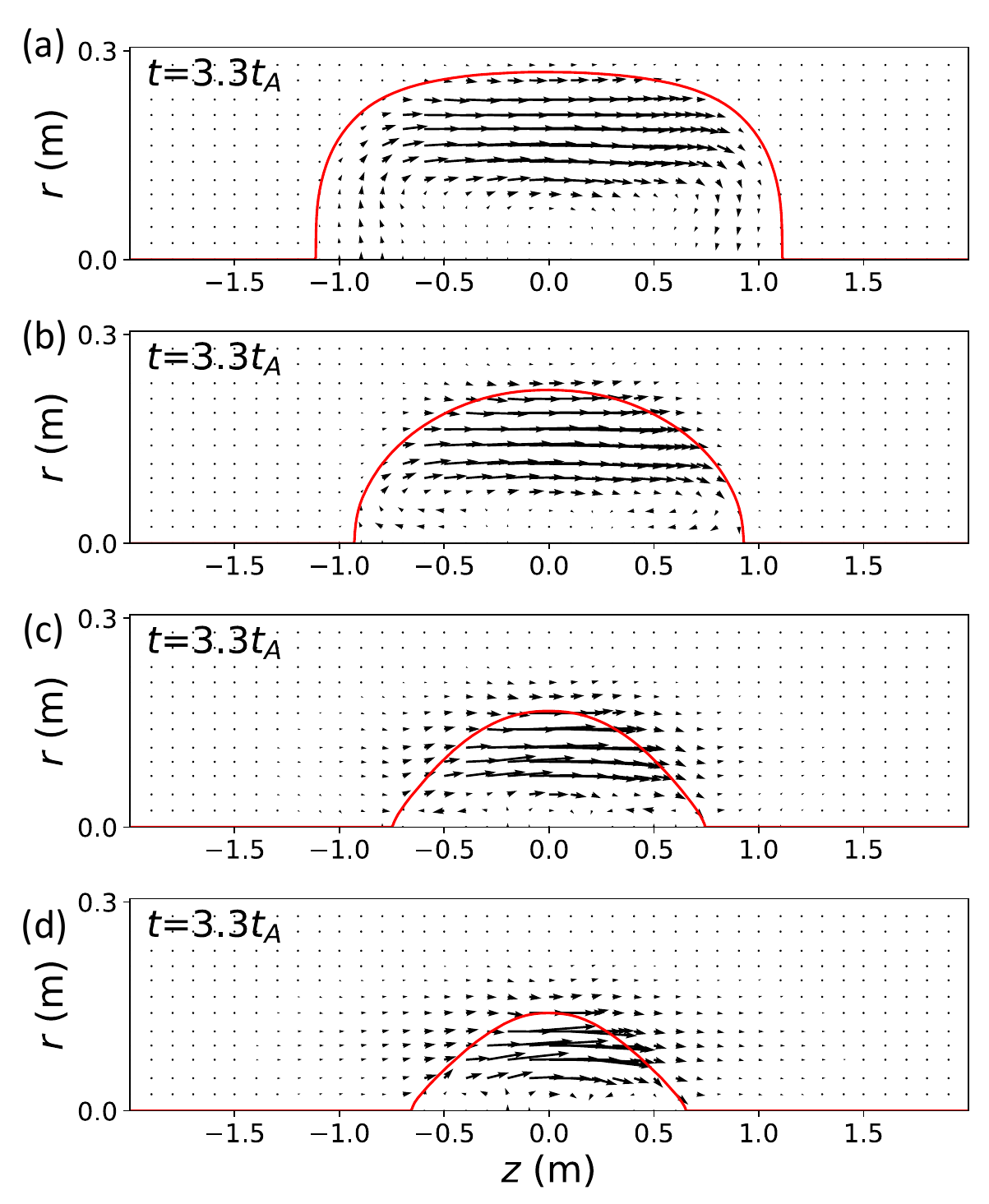}
  \caption{Velocity vectors of the $n=1$ tilt mode and the separatrix locations for the cases with the compression field ramping rates of (a) $0.03B_{w0}/\mathrm{\mu s}$, (b) $0.09B_{w0}/\mathrm{\mu s}$, (c) $0.18B_{w0}/\mathrm{\mu s}$ and (d) $0.24 B_{w0}/\mathrm{\mu s}$.}
  \label{fig: rpvvplot}
\end{figure*}

For higher ramping rate, the FRC contracts faster and evolves to a more oblate shape with a reduced plasma volume (Figure~\ref{fig: rpstream_z0uz_comp}). This rapid contraction is consistent with the observed increase in the growth rate of the tilt mode and the corresponding reduction in nonlinear saturation energy at higher compression field ramping rates. For the compression field ramping rate of  $0.03B_{w0}/\mathrm{\mu s}$, the plasma temperature stops increasing when the magnetic compression ratio $B_{w2}/B_{w1}$ exceeds 1.6, whereas for ramping rate of $0.24B_{w0}/\mathrm{\mu s}$ the decrease of the temperature starts at a higher compression ratio of about $B_{w2}/B_{w1}\sim3.6$. Although a higher compression field ramping rate allows reaching  larger compression ratios before the peak compressional heating, the improvement is limited and the enhanced compression dynamics does not stabilize the tilt mode. Since the tilt mode grows on the  Alfv\'{e}nic timescale and even becomes more unstable with increasing ramping rates, it may be difficult to prevent a tilt disruption up to a moderate peak compression ratio like $B_{w2}/B_{w1}\sim6-7$~\cite{rej1992high} by merely increasing the compression field ramping rate alone. During the early linear growth phase of the tilt mode, the $n=1$ displacement outside the separatrix becomes more pronounced as the ramping rate increases 
 (Figure~\ref{fig: rpvvplot}). Thus, the tilt mode may become easier to identify experimentally in the early growth phase for more dynamic compression processes. Nevertheless, figures~\ref{fig: e1modestruct} and~\ref{fig: rpstream_z0uz_comp} indicate that the mode structure in the $r$-$\theta$ plane does not vary significantly across different compression field ramping rates.

\subsection{Effects of toroidal flow}

FRCs can exhibit strong toroidal flow due to the spontaneous spin-up or external driving methods such as rotating magnetic fields~\cite{steinhauer2006modeling,ren2021plasma,ren2022improvement}. Earlier studies~\cite{belova2000numerical, milroy1989nonlinear, horiuchi1989full, clemente1983tilting, ishida1988variational} investigated the tilt instability of FRCs in presence of steady-state toroidal rotation. Their results indicate that toroidal rotation provides negligible stabilizing effects on the tilt mode at moderate Mach numbers ($M_A<0.5$), where  $M_A=\Omega_{0m}/\Omega_A=V_{0m}/V_A$, $\Omega_A=V_A/r$ and $V_{0m}$ is the initial maximum magnitude of the toroidal velocity $V_t$. The toroidal velocity $V_t=-u_\theta$ is oriented in the co-rotating direction opposite to the positive $\theta$ direction. Although the growth rate of the tilt mode can be significantly reduced at higher toroidal rotation rates ($M_A>1$)~\cite{ belova2000numerical,milroy1989nonlinear,ishida1988variational}, such high rotation rates are generally considered unrealistic. Here we examine the toroial flow effects on the tilt mode at a smaller rotation rate ($M_A\sim0.5$) during the dynamic compression process and evaluates whether the dynamic compression can enhance the stabilizing effects of toroidal rotation.

\begin{figure*}[!htbp]
  \centering
  \includegraphics[width=0.7\textwidth]{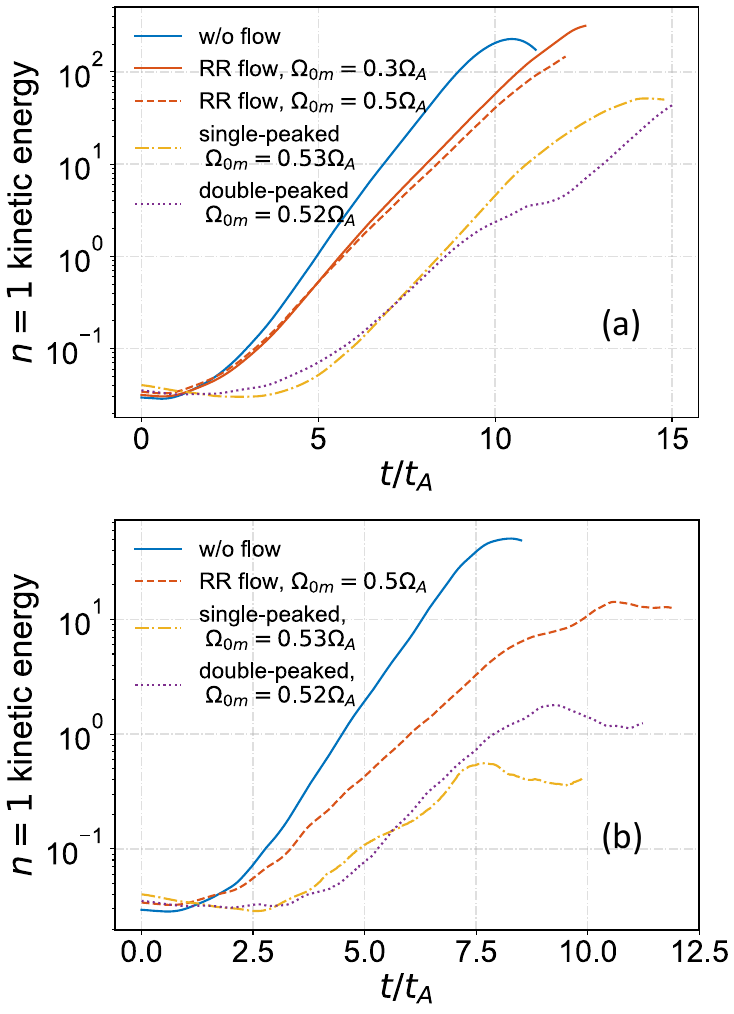}
  \caption{The temporal evolution of the $n=1$ kinetic energy in the simulation cases with various maximum initial rotation magnitudes and rotation profiles, and the compression field ramping rates of (a) $0.12B_{w0}/\mathrm{\mu s}$ and (b) $0.24B_{w0}/\mathrm{\mu s}$.}
  \label{fig: flowencomp_0.12_0.24bw}
\end{figure*}

\begin{figure*}[!htbp]
  \centering
  \includegraphics[width=0.7\textwidth]{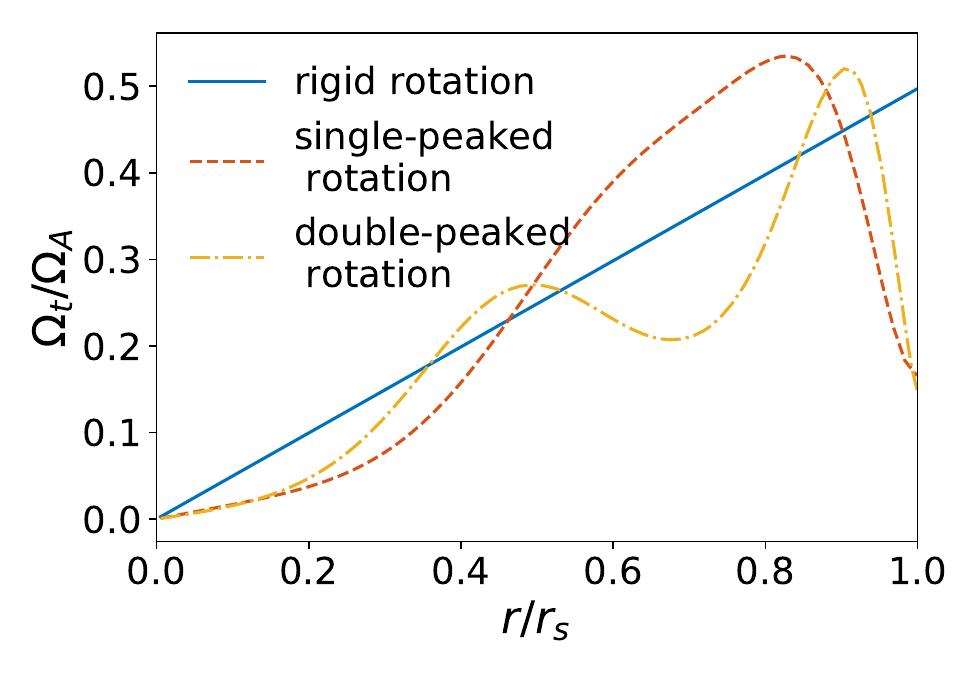}
  \caption{The initial rigid, single-peaked and double-peaked toroidal rotation profiles in the $z=0$ middle plane.}
  \label{fig: flow1dprofile}
\end{figure*}

\begin{figure*}[!htbp]
  \centering
  \includegraphics[width=0.5\textwidth]{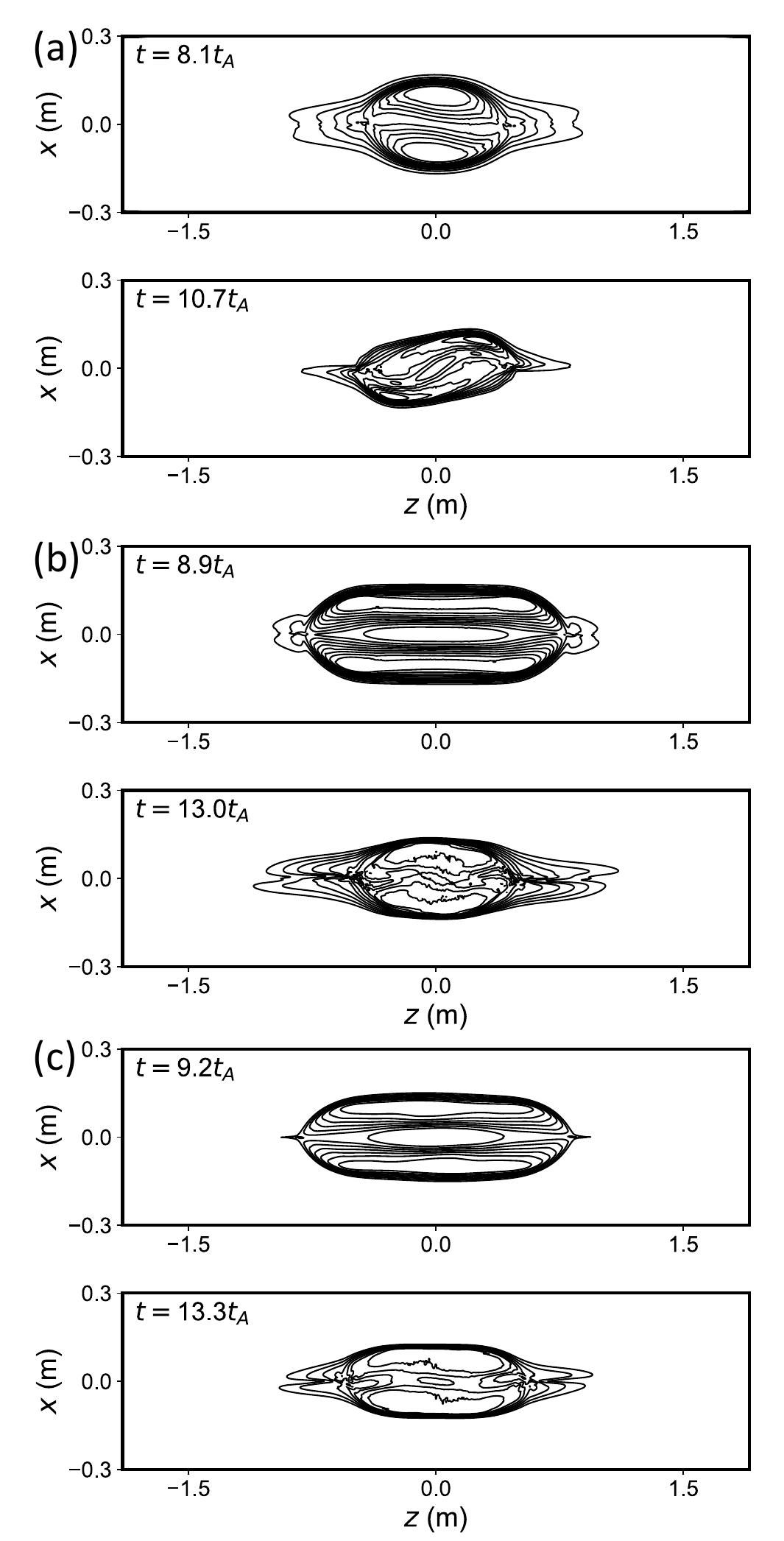}
  \caption{The pressure contours at different times during compression in the presence of the (a) initial RR flow and $\Omega_{0m}=0.5\Omega_A$, (b) initial single-peaked rotation and $\Omega_{0m}=0.53\Omega_A$ and (c) initial double-peaked rotation and $\Omega_{0m}=0.52\Omega_A$. The compression field ramping rate is $0.12B_{w0}/\mathrm{\mu s}$.}
  \label{fig: flowpcontour0.12bw0}
\end{figure*}

\begin{figure*}[!htbp]
  \centering
  \includegraphics[width=1.0\textwidth]{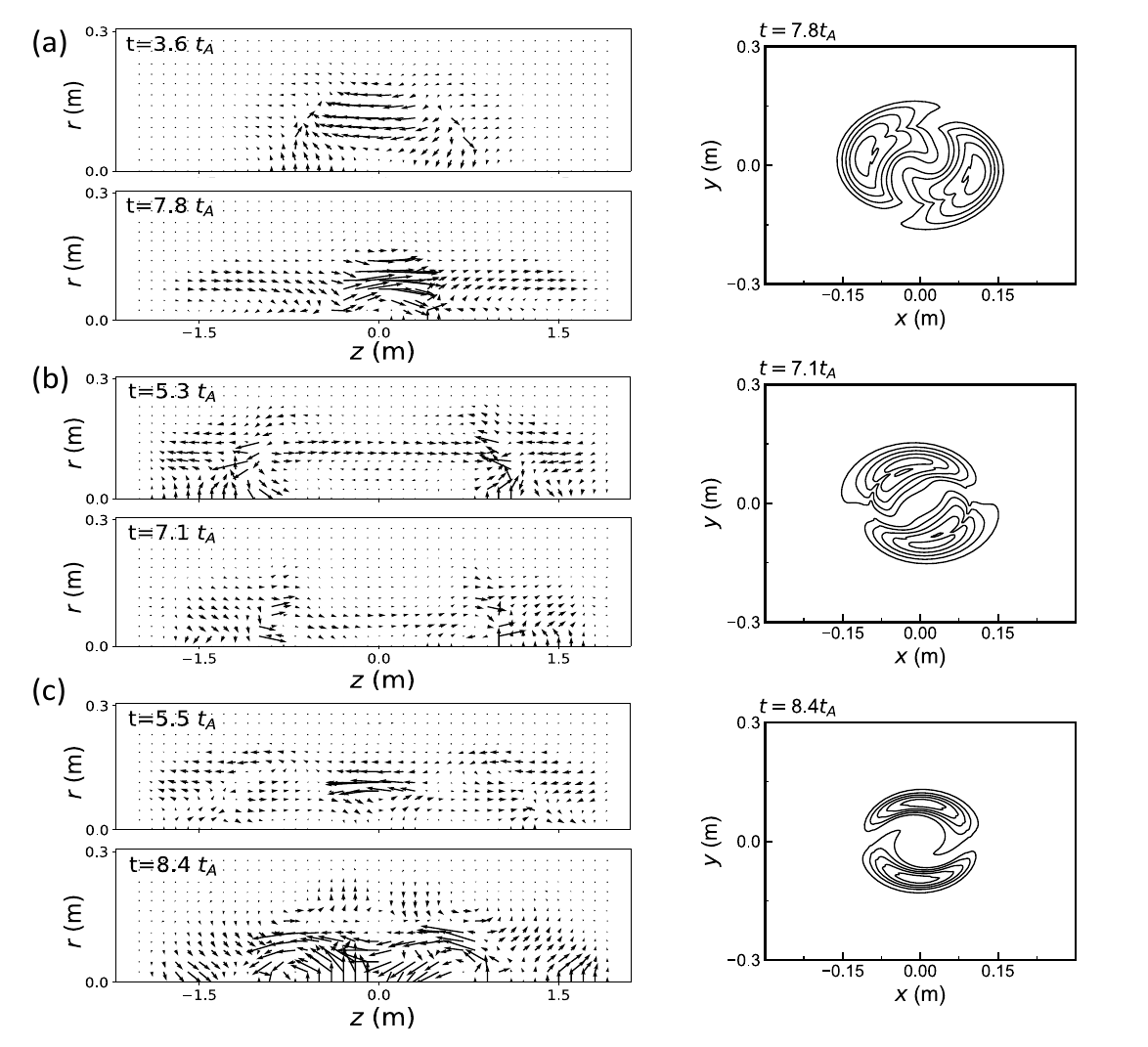}
  \caption{The velocity vectors in the poloidal plane (left) and axial velocity contours in the $z=0$ middle plane (right) of the $n=1$ tilt mode at different times during compression for the cases with (a) initial RR flow and $\Omega_{0m}=0.5\Omega_A$, (b) initial single-peaked rotation and $\Omega_{0m}=0.53\Omega_A$ and (c) initial double-peaked rotation and $\Omega_{0m}=0.52\Omega_A$. The compression field ramping rate is $0.12B_{w0}/\mathrm{\mu s}$ and corresponds to the cases shown in figure~\ref{fig: flowpcontour0.12bw0}.}
  \label{fig: flowvv_z0uz_0.12bw0}
\end{figure*}

Figure~\ref{fig: flowencomp_0.12_0.24bw}~(a) compares the $n=1$ kinetic energy evolution for cases with and without initial toroidal rotation under a compression field ramping rate of $0.12B_{w0}/\mathrm{\mu s}$. The corresponding equilibrium rotation profiles are shown in figure~\ref{fig: flow1dprofile}. The results indicate that the toroidal rotation reduces both the growth rate and the nonlinear saturation amplitude of the tilt mode energy. The initial RR flow exhibits the weakest stabilizing effects due to its lower flow shear among the rotation profiles. Moreover, higher initial rotation rates lead to stronger stabilizing effects. Figure~\ref{fig: flowpcontour0.12bw0} illustrates the pressure contours for each case. Compared to the standard case without initial flow discussed in section~\ref{sec: typical case}, the presence of initial toroidal rotation delays the onset of significant tilt deformation by $1$-$6t_A$ for various initial rotation profiles. In addition, the toroidal flow alters the pressure distribution through modifications in the initial equilibrium as described in equations~(\ref{eqn: gs})-(\ref{eqn: rho wi flow}). The tilt mode structure also exhibits significant changes in the presence of the initial toroidal rotation. The tilt mode evolves in both the poloidal and $r$-$\theta$ planes, as shown in figure~\ref{fig: flowvv_z0uz_0.12bw0}. The mode structure rotates toroidally and exhibits more displacement outside the separatrix compared to the case without initial toroidal rotation.

\begin{figure*}[!htbp]
  \centering
  \includegraphics[width=0.8\textwidth]{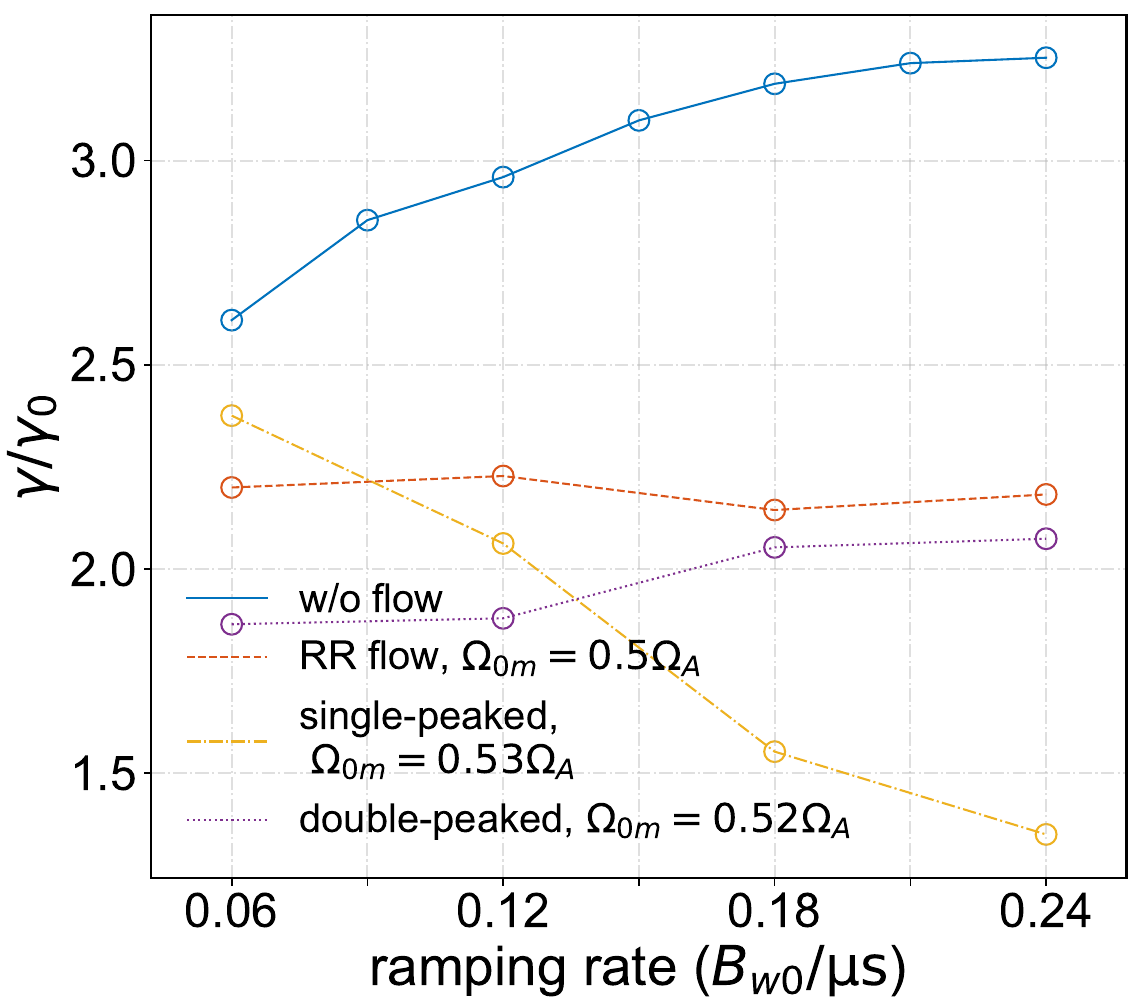}
  \caption{Variations of the tilt mode growth rate with the compression field ramping rate for simulation cases with various initial toroidal rotation profiles.}
  \label{fig: flowgr}
\end{figure*}

The stabilizing effects of toroidal rotation on the tilt mode are observed to become more pronounced with increasing ramping rates of the compression magnetic field. As shown in figures~\ref{fig: flowencomp_0.12_0.24bw} and~\ref{fig: flowgr}, cases with  RR and double-peaked initial toroidal rotation show no longer significant increase, and the single-peaked case demonstrates even substantial decrease in the growth rate as the compression field ramping rate increases, in contrast to the monotonic increase in growth rate in cases without initial toroidal rotation.  The nonlinear saturation amplitude of the $n=1$ kinetic energy is further reduced at higher compression field ramping rates when the initial toroidal rotation is present. For the simulation case with an initially double-peaked toroidal rotation profile and $M_A\sim 0.52$, the compressional heating terminates at a magnetic compression ratio of $B_{w2}/B_{w1}\sim5.3$ under the compression field ramping rate of $0.24B_{w0}/\mathrm{\mu s}$, which is close to the moderate compression ratio of $6-7$ achieved in experiments~\cite{rej1992high}.

\begin{figure*}[!htbp]
  \centering
  \includegraphics[width=0.7\textwidth]{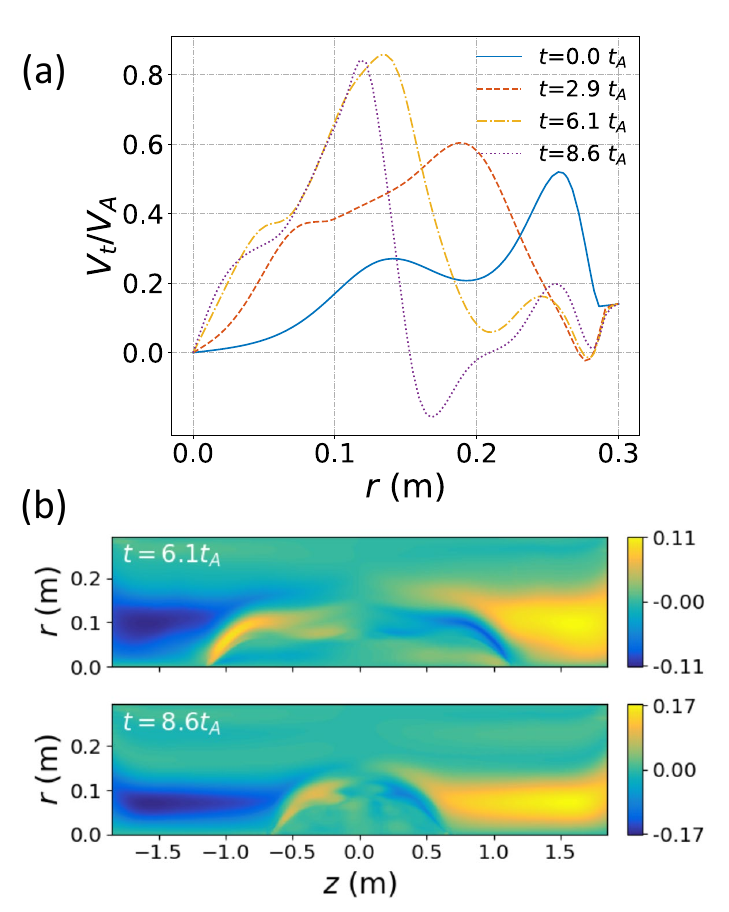}
  \caption{ (a) Radial profiles of toroidal velocity along the $z=0$ middle plane and (b) contours of $B_\theta$ at different times during compression for the case with initial double-peaked rotation profile and $\Omega_{0m}\sim0.52\Omega_{A}$, and the compression field ramping rate of $0.18B_{w0}/\mathrm{\mu s}$.}
  \label{fig: flowvt1d_double0.18bw0}
\end{figure*}

\begin{figure*}[!htbp]
  \centering
  \includegraphics[width=0.8\textwidth]{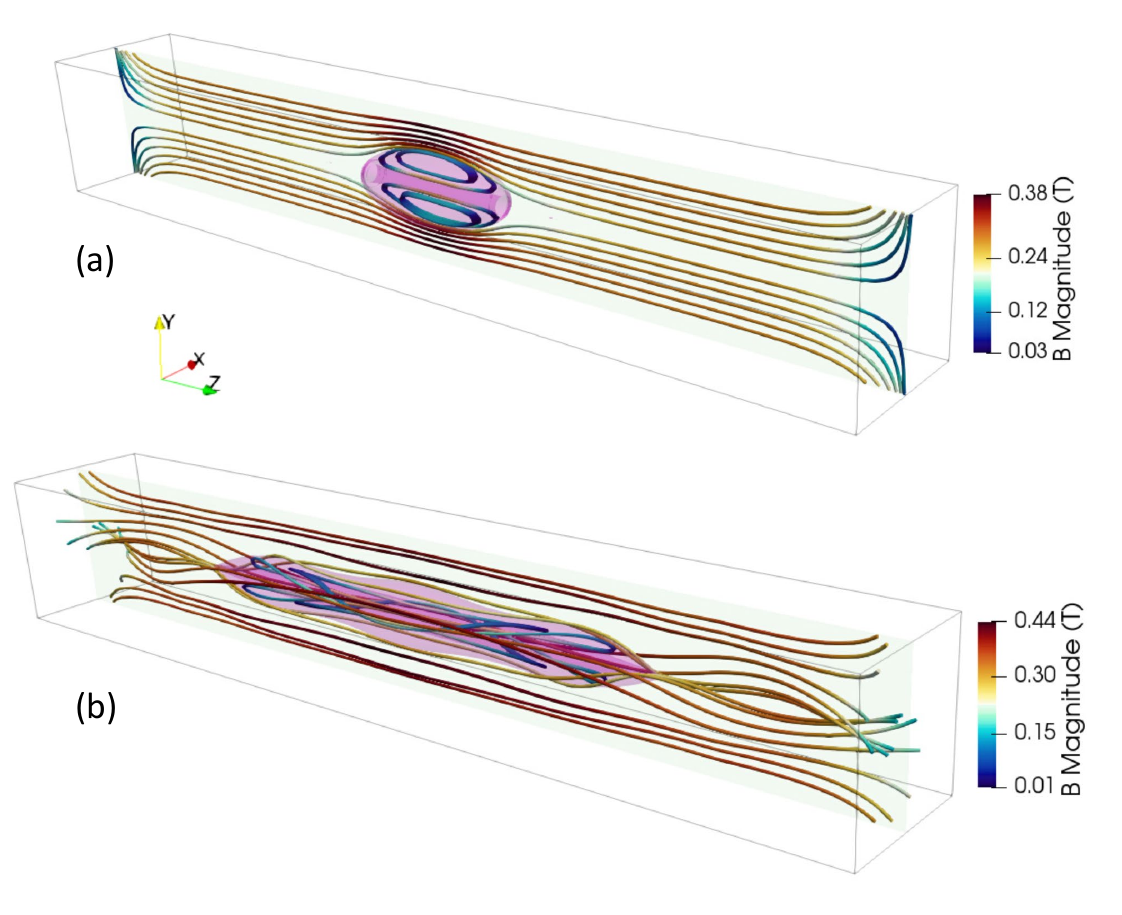}
  \caption{The 3D pressure contours and magnetic field lines for the cases (a) without initial toroidal flow at $5.2t_A$ and (b) with initially double-peaked rotation profile ($\Omega_{0m}=0.52\Omega_A$) at $6.1t_A$. The compression field ramping rate is $0.18B_{w0}/\mathrm{\mu s}$.}
  \label{fig: flow3dstream0.18bw0}
\end{figure*}

The enhanced stabilizing effects of toroidal rotation on the tilt mode at higher compression field ramping rates may be understood as a natural consequence of deviation from the initial equilibrium state of a rotating FRC. In the initial equilibrium with toroidal rotation $\Omega(\psi)$, the flow velocity is $\vec{u}=u_\theta \hat{e}_\theta=r^2\Omega\nabla\theta$ and the equilibrium magnetic field is given by $\vec{B} = \nabla \psi \times \nabla \theta$. Then the toroidal component of the magnetic field evolves as
\begin{equation}
    \label{eqn: bth}
     \frac{\partial B_\theta}{\partial t}   = \hat{e}_\theta \cdot (\nabla \Omega \times \nabla \psi),
\end{equation}
if the plasma resistivity is neglected~\cite{guzdar1985role, wira1990toroidal}. Equation~\ref{eqn: bth} shows that a toroidal magnetic field is generated when the rotation frequency $\Omega$ is no longer a flux function during the dynamic compression. Because $\Omega$ and $\psi$ are symmetric about the $z=0$ middle plane, the resulting $B_\theta$ is antisymmetric with respect to $z=0$. This antisymmetric $B_\theta$ can then induce an electromagnetic torque on a cylindrical region of radius $r_{end}$ that encloses the FRC, where
\begin{equation}
   \tau_z = \int_0^{r_{end}} \frac{2\pi}{\mu_0} r^2 B_\theta B_z dr 
\end{equation}
is the axial electromagnetic torque from each end surface of the cylinder~\cite{pustovitov2011integral, milroy2010extended}. This electromagnetic torque decelerates the rotation outside the separatrix and leads to a pronounced flow shear, as also demonstrated in the 2D magnetic compression simulations~\cite{ma2024invariant}. Figure~\ref{fig: flowvt1d_double0.18bw0} shows the contours of the toroidal magnetic field and the significant flow shear during the compression process. Both features have stabilizing effects and grow in significance as magnetic compression becomes more dynamic. In the presence of initial toroidal rotation, the FRC exhibits a more complex field line structure during compression compared to the case without initial toroidal flow, as shown in figure~\ref{fig: flow3dstream0.18bw0}.

\section{Discussion and conclusions}
\label{sec: Discussion and conclusions}

The nonlinear MHD simulations of the tilt instability during the dynamic FRC magnetic compression process have been conducted using the NIMROD code. The influences of the compression field ramping rate and toroidal flow on the tilt mode evolution are also examined.

The tilt instability grows on the Alfv\'{e}n timescale and disrupts the FRC in the early stage of the magnetic compression process. The tilt mode causes notable deformation during the linear growth phase, which is followed by a complete disruption of the FRC in the nonlinear phase. There is no evidence of dynamic nonlinear stabilization. With an increased compression field ramping rate, the FRC shrinks rapidly into an oblate shape with reduced plasma volume. The linear growth rate of the tilt mode increases with the compression field ramping rate and approaches an asymptotic value, whereas the nonlinear saturation magnitude of the $n=1$ kinetic energy decreases. Enhanced compression also induces larger velocity perturbation outside the separatrix. Although a more rapidly ramping of compression field can lead to higher compression ratios before the peak compressional heating, such a scheme has limited enhancement in compression ratio and is not effective for practical compressional heating.

Toroidal flow can reduce both the linear growth rate and the nonlinear saturation amplitude of the tilt mode. The mode structure rotates toroidally and shows increased velocity perturbation outside the separatrix in comparison to the case without initial toroidal rotation. Higher initial rotation rates produce stronger stabilizing effects. In addition, the stabilizing effects of the toroidal rotation become more significant with higher ramping rate of the compression field. As the compression process becomes more dynamic, the induced toroidal magnetic field and strong flow shear contribute to additional stabilizing effects. In the simulation with an initial double-peaked rotation profile and $M_A\sim0.52$, the compressional heating terminates at a magnetic compression ratio of $B_{w2}/B_{w1}\sim5.3$ that is close to the moderate compression ratio of 6-7 found in experiments~\cite{rej1992high}. While neither rapid compression nor toroidal rotation  suppresses the tilt instability completely during compression, their combined effects extend the plasma lifetime and enable higher effective magnetic compression ratios.

The high-$\bar{s}$ operation is excepted to substantially improve the reactor potential of FRCs, particularly when the advanced fuels are considered~\cite{hill2008report, slough1995transport, slough1992confinement, belova2001numerical}. The active stabilizing methods that are effective for steady-state FRCs, such as the neutral beam injection, may lose their efficacy during the compression process due to the short pulse duration and rapid changes in the FRC shape. On the other hand, a complete stabilization of the tilt mode may not be required for the pulsed systems, provided that the significant tilt deformation can remain suppressed until the time of the peak compression. These considerations highlight the importance of passive or spontaneous stabilizing mechanisms during compression. This study demonstrates that the stabilizing effects of the toroidal flow increase under more dynamic conditions. Moreover, additional stabilizing effects may also contribute to the tilt stabilization. For example, the kinetic effects have been shown to modify the tilt mode structure and extend FRC lifetime even in the high-$\bar{s}$ regime~\cite{belova2000numerical,belova2004kinetic, belova2003kinetic}. Considering all potential spontaneous stabilizing mechanisms, it is possible that the compressional heating can take place in the high-$\bar{s}$ regime without tilt disruption within the time duration of a pulse in the presence of initial toroidal flow. Future studies plan to incorporate kinetic effects into the simulations in order to further evaluate such a scenario.

\section{Acknowledgement}

This work was supported by the National MCF Energy R\&D Program of China (Grant No. 2019YFE03050004), the National Natural Science Foundation of China (Nos. 62201217 and 51821005), the Wuhan Municipal Science and Technology Innovation Platform Program (Grant No. 20200766), the HUST Fundamental Research Support Program (Grant No. 2023BR010), the U.S. Department of Energy (Grant No. DE-FG02-86ER53218) and the Hubei International Science and Technology Cooperation Project under Grant No. 2022EHB003. The computing work in this paper was supported by the Public Service Platform of High Performance Computing by Network and Computing Center of HUST. The authors are very grateful for the supports from the NIMROD team and the J-TEXT team.

\section*{References}
\normalsize
\bibliography{references}

\providecommand{\newblock}{}
\begin{thebibliography}{10}
\expandafter\ifx\csname url\endcsname\relax
  \def\url#1{{\tt #1}}\fi
\expandafter\ifx\csname urlprefix\endcsname\relax\def\urlprefix{URL }\fi
\providecommand{\eprint}[2][]{\url{#2}}

\bibitem{tuszewski1988field}
Tuszewski M. 1988 Field reversed configurations {\em Nucl. Fusion\/} {\bf 28}
  2033

\bibitem{rej1986experimental}
Rej D.J. {\em et~al\/} 1986 Experimental studies of field-reversed
  configuration translation {\em Phys. Fluids\/} {\bf 29} 852--862

\bibitem{steinhauer1996frc}
Steinhauer L.C. 1996 {FRC} 2001: a white paper on {FRC} development in the next
  five years {\em Fus. Technol.\/} {\bf 30} 116

\bibitem{liao2022first}
Liao H. {\em et~al\/} 2022 First direct experimental evidence of the merging of
  two colliding field reversed configurations {\em Plasma Phys. Control.
  Fusion\/} {\bf 64} 115003

\bibitem{liao2022dynamic}
Liao H. {\em et~al\/} 2022 The dynamic axial compression of {FRC} with
  high-speed translated $\theta$-pinch plasma {\em Plasma Phys. Control.
  Fusion\/} {\bf 64} 105015

\bibitem{sekiguchi2018super}
Sekiguchi J. {\em et~al\/} 2018 Super-{A}lfv{\'e}nic translation of a
  field-reversed configuration into a large-bore dielectric chamber {\em Rev.
  Sci. Instrum.\/} {\bf 89} 013506

\bibitem{kobayashi2021experimental}
Kobayashi D. and Asai T. 2021 Experimental evidence for super-{A}lfv{\'e}nic
  acceleration of the field-reversed configuration due to a magnetic pressure
  gradient {\em Phys. Plasmas\/} {\bf 28} 022101

\bibitem{asai2021observation}
Asai T. {\em et~al\/} 2021 Observation of self-organized {FRC} formation in a
  collisional-merging experiment {\em Nucl. Fusion\/} {\bf 61} 096032

\bibitem{asai2019collisional}
Asai T. {\em et~al\/} 2019 Collisional merging formation of a field-reversed
  configuration in the {FAT-CM} device {\em Nucl. Fusion\/} {\bf 59} 056024

\bibitem{steinhauer2011review}
Steinhauer L.C. 2011 Review of field-reversed configurations {\em Phys.
  Plasmas\/} {\bf 18} 070501

\bibitem{gota2024enhanced}
Gota H. {\em et~al\/} 2024 Enhanced plasma performance in {C-2W} advanced
  beam-driven field-reversed configuration experiments {\em Nucl. Fusion\/}
  {\bf 64} 112014

\bibitem{kirtley2023fundamental}
Kirtley D. and Milroy R. 2023 Fundamental scaling of adiabatic compression of
  field reversed configuration thermonuclear fusion plasmas {\em J. Fusion
  Energy\/} {\bf 42} 30

\bibitem{spencer1983adiabatic}
Spencer R.L. {\em et~al\/} 1983 Adiabatic compression of elongated
  field-reversed configurations {\em Phys. Fluids\/} {\bf 26} 1564

\bibitem{woodruff2008adiabatic}
Woodruff S. {\em et~al\/} 2008 Adiabatic compression of a doublet field
  reversed configuration ({FRC}) {\em J. Fusion Energy\/} {\bf 27} 128

\bibitem{ma2023mhd}
Ma Y. {\em et~al\/} 2023 Mhd simulation on magnetic compression of field
  reversed configurations with {NIMROD} {\em Nucl. Fusion\/} {\bf 63} 046017

\bibitem{degnan2013recent}
Degnan J.H. {\em et~al\/} 2013 Recent magneto-inertial fusion experiments on
  the field reversed configuration heating experiment {\em Nucl. Fusion\/} {\bf
  53} 093003

\bibitem{okada2001experiments}
Okada S. {\em et~al\/} 2001 Experiments on additional heating of {FRC} plasmas
  {\em Nucl. Fusion\/} {\bf 41} 625

\bibitem{gota2021overview}
Gota H. {\em et~al\/} 2021 Overview of {C-2W}: high temperature, steady-state
  beam-driven field-reversed configuration plasmas {\em Nucl. Fusion\/} {\bf
  61} 106039

\bibitem{galea2023princeton}
Galea C. {\em et~al\/} 2023 The princeton field-reversed configuration for
  compact nuclear fusion power plants {\em J. Fusion Energy\/} {\bf 42} 4

\bibitem{hoffman2009advances}
Hoffman A.L. {\em et~al\/} 2009 Advances in singly connected closed field line
  plasma devices and extrapolation to {POP} level experiments and reactors {\em
  Nucl. Fusion\/} {\bf 49} 055018

\bibitem{ono2024ion}
Ono Y, Tanabe H and Inomoto M 2024 Ion heating characteristics of merging
  spherical tokamak plasmas for burning high-beta plasma formation {\em Nucl.
  Fusion\/} {\bf 64} 086020

\bibitem{rej1992high}
Rej D.J. {\em et~al\/} 1992 High-power magnetic-compression heating of
  field-reversed configurations {\em Phys. Fluids B: Plasma Phys.\/} {\bf 4}
  1909

\bibitem{slough2011creation}
Slough J. {\em et~al\/} 2011 Creation of a high-temperature plasma through
  merging and compression of supersonic field reversed configuration plasmoids
  {\em Nucl. Fusion\/} {\bf 51} 053008

\bibitem{turchi2025progress}
Turchi P.J. 2025 Progress and issues with pulsed magnetic fusion {\em Phys.
  Plasmas\/} {\bf 32}

\bibitem{hill2008report}
Hill D.N. and Hazeltine R. 2008 Report of the {FESAC} toroidal alternates panel
  {\em U.S. Department of Energy, Washington DC\/}

\bibitem{rosenbluth1979mhd}
Rosenbluth M.N. and Bussac M.N. 1979 {MHD} stability of spheromak {\em Nucl.
  Fusion\/} {\bf 19} 489

\bibitem{clemente1981tilting}
Clemente R.A. and Milovich J.L. 1981 The tilting mode in field-reversed
  configurations {\em Phys. Lett. A\/} {\bf 85} 148--150

\bibitem{hammer1981mhd}
Hammer J.H. 1981 {MHD} tilting modes for nearly spherical compact toroids with
  arbitrary plasma pressure {\em Nucl. Fusion\/} {\bf 21} 488

\bibitem{bellan2000spheromaks}
Bellan P.M. 2000 {\em Spheromaks: a practical application of
  magnetohydrodynamic dynamos and plasma self-organization\/} (World
  Scientific)

\bibitem{belova2000numerical}
Belova E.V. {\em et~al\/} 2000 Numerical study of tilt stability of prolate
  field-reversed configurations {\em Phys. Plasmas\/} {\bf 7} 4996--5006

\bibitem{milroy1989nonlinear}
Milroy R.D. {\em et~al\/} 1989 Nonlinear magnetohydrodynamic studies of the
  tilt mode in field-reversed configurations {\em Phys. Fluids B: Plasma
  Phys.\/} {\bf 1} 1225--1232

\bibitem{schwarzmeier1983magnetohydrodynamic}
Schwarzmeier J.L. {\em et~al\/} 1983 Magnetohydrodynamic equilibrium and
  stability of field-reversed configurations {\em Phys. Fluids\/} {\bf 26} 1295

\bibitem{cobb1993profile}
Cobb J.W. {\em et~al\/} 1993 Profile stabilization of tilt mode in a
  field-reversed configuration {\em Phys. Fluids B: Plasma Phys.\/} {\bf 5}
  3227--3238

\bibitem{iwasawa2000ideal}
Iwasawa N. {\em et~al\/} 2000 Ideal magnetohydrodynamic stability of static
  field reversed configurations {\em J. Phys. Soc. Jpn.\/} {\bf 69} 451--463

\bibitem{horiuchi1989full}
Horiuchi R. and Sato T. 1989 Full magnetohydrodynamic simulation of the tilting
  instability in a field-reversed configuration {\em Phys. Fluids B\/} {\bf 1}
  pp--581

\bibitem{nishimura1997tilt}
Nishimura K. {\em et~al\/} 1997 Tilt stabilization by cycling ions crossing
  magnetic separatrix in a field-reversed configuration {\em Phys. Plasmas\/}
  {\bf 4} 4035--4042

\bibitem{barnes1986kinetic}
Barnes D.C. {\em et~al\/} 1986 Kinetic tilting stability of field-reversed
  configurations {\em Phys. Fluids\/} {\bf 29} 2616--2629

\bibitem{belova2004kinetic}
Belova E.V. {\em et~al\/} 2004 Kinetic effects on the stability properties of
  field-reversed configurations. {II}. {N}onlinear evolution {\em Phys.
  Plasmas\/} {\bf 11} 2523--2531

\bibitem{slough1995transport}
Slough J.T. {\em et~al\/} 1995 Transport, energy balance, and stability of a
  large field-reversed configuration {\em Phys. Plasmas\/} {\bf 2} 2286--2291

\bibitem{slough1992confinement}
Slough J.T. {\em et~al\/} 1992 Confinement and stability of plasmas in a
  field-reversed configuration {\em Phys. Rev. Lett.\/} {\bf 69} 2212

\bibitem{belova2001numerical}
Belova E.V. {\em et~al\/} 2001 Numerical study of global stability of oblate
  field-reversed configurations {\em Phys. Plasmas\/} {\bf 8} 1267--1277

\bibitem{slough1993stability}
Slough J.T. and Hoffman A.L. 1993 Stability of field-reversed configurations in
  the large s experiment ({LSX}) {\em Phys. Fluids B: Plasma Phys.\/} {\bf 5}
  4366--4377

\bibitem{sovinec2004nonlinear}
Sovinec C.R. {\em et~al\/} 2004 Nonlinear magnetohydrodynamics simulation using
  high-order finite elements {\em J. Comput. Phys.\/} {\bf 195} 355

\bibitem{li2021solving}
Li H. and Zhu P. 2021 Solving the {Grad--Shafranov} equation using spectral
  elements for tokamak equilibrium with toroidal rotation {\em Comput. Phys.
  Commun.\/} {\bf 260} 107264

\bibitem{howell2014solving}
Howell E.C. and Sovinec C.R. 2014 Solving the {Grad--Shafranov} equation with
  spectral elements {\em Comput. Phys. Commun.\/} {\bf 185} 1415

\bibitem{evangelias2016axisymmetric}
Evangelias A. and Throumoulopoulos G.N. 2016 Axisymmetric equilibria with
  pressure anisotropy and plasma flow {\em Plasma Phys. Control. Fusion\/} {\bf
  58} 045022

\bibitem{guazzotto2009magnetic}
Guazzotto L. and Paccagnella R. 2009 Magnetic field profiles in fusion plasmas
  in the presence of equilibrium flow {\em Plasma Phys. Control. Fusion\/} {\bf
  51} 065013

\bibitem{maschke1980exact}
Maschke E.K. and Perrin H. 1980 Exact solutions of the stationary {MHD}
  equations for a rotating toroidal plasma {\em Plasma Phys.\/} {\bf 22} 579

\bibitem{sen1997tokamak}
Sen A. and Ramos J.J. 1997 Tokamak equilibria and stability with arbitrary
  flows {\em Plasma Phys. Control. Fusion\/} {\bf 39} A323

\bibitem{li2021formation}
Li H. and Zhu P. 2021 Formation of edge pressure pedestal and reversed magnetic
  shear due to toroidal rotation in a tokamak equilibrium {\em Phys. Plasmas\/}
  {\bf 28} 054505

\bibitem{ono2003spontaneous}
Ono Y. {\em et~al\/} 2003 Spontaneous and artificial generation of sheared-flow
  in oblate {FRC}s in {TS}-3 and 4 {FRC} experiments {\em Nucl. Fusion\/} {\bf
  43} 649

\bibitem{barnes1991stabilization}
Barnes D.C. and Milroy R.D. 1991 Stabilization of the field-reversed
  configuration ({FRC}) tilt instability with energetic ion beams {\em Phys.
  Fluids B: Plasma Phys.\/} {\bf 3} 2609--2616

\bibitem{steinhauer2006modeling}
Steinhauer L.C. {\em et~al\/} 2006 Modeling of field-reversed configuration
  experiment with large safety factor {\em Phys. Plasmas\/} {\bf 13}

\bibitem{ren2021plasma}
Ren B. {\em et~al\/} 2021 Plasma rotation driven by rotating magnetic fields
  {\em Plasma Phys. Control. Fusion\/} {\bf 63} 035027

\bibitem{ren2022improvement}
Ren B. {\em et~al\/} 2022 Improvement of radial confinement of plasma via
  applying rotating magnetic fields {\em Plasma Phys. Control. Fusion\/} {\bf
  64} 095016

\bibitem{clemente1983tilting}
Clemente R.A. and Milovich J.L. 1983 The tilting mode in rigidly rotating
  field-reversed configurations {\em Phys. Fluids\/} {\bf 26} 1874--1876

\bibitem{ishida1988variational}
Ishida A. {\em et~al\/} 1988 Variational formulation for a multifluid flowing
  plasma with application to the internal tilt mode of a field-reversed
  configuration {\em Phys. Fluids\/} {\bf 31} 3024--3034

\bibitem{guzdar1985role}
Guzdar P.N. {\em et~al\/} 1985 The role of magnetic reconnection and
  differential rotation in spheromak formation {\em Phys. Fluids\/} {\bf 28}
  3154--3166

\bibitem{wira1990toroidal}
Wira K. and Pietrzyk Z.A. 1990 Toroidal field generation and magnetic field
  relaxation in a conical-theta-pinch-generated configuration {\em Phys. Fluids
  B: Plasma Phys.\/} {\bf 2} 561--573

\bibitem{pustovitov2011integral}
Pustovitov V.D. 2011 Integral torque balance in tokamaks {\em Nucl. Fusion\/}
  {\bf 51} 013006

\bibitem{milroy2010extended}
Milroy R.D. {\em et~al\/} 2010 Extended magnetohydrodynamic simulations of
  field reversed configuration formation and sustainment with rotating magnetic
  field current drive {\em Phys. Plasmas\/} {\bf 17} 062502

\bibitem{ma2024invariant}
Ma Y. {\em et~al\/} 2024 Invariant regimes of spencer scaling law for magnetic
  compression of rotating {FRC} plasma {\em Nucl. Fusion\/} {\bf 64} 126024

\bibitem{belova2003kinetic}
Belova E.V. {\em et~al\/} 2003 Kinetic effects on the stability properties of
  field-reversed configurations. {I}. {L}inear stability {\em Phys. Plasmas\/}
  {\bf 10} 2361--2371

\end{thebibliography}
\clearpage


\clearpage

\clearpage

\clearpage

\clearpage

\clearpage

\clearpage

\clearpage

\clearpage

\clearpage

\clearpage

\clearpage

\clearpage

\clearpage

\end{document}